# Efficient Electrical Detection of Mid-Infrared Graphene Plasmons at Room Temperature

Qiushi Guo[1]†, Renwen Yu[2]†, Cheng Li[1], Shaofan Yuan[1], Bingchen Deng[1], F. Javier García de Abajo[2, 3]* and Fengnian Xia[1]*

[1]Department of Electrical Engineering, Yale University, New Haven, Connecticut 06511, USA

[2]ICFO-Institut de Ciencies Fotoniques, The Barcelona Institute of Science and Technology, 08860 Castelldefels, Barcelona, Spain.

[3]Institució Catalana de Recerca i Estudis Avançats (ICREA), Passeig Lluís Companys 23, 08010 Barcelona, Spain

*Correspondence to: fengnian.xia@yale.edu; javier.garciadeabajo@nanophotonics.es.

† These authors contributed equally to this work.

**Abstract**

Optical excitation and subsequent decay of graphene plasmons can produce a significant increase in charge-carrier temperature. An efficient method to convert this temperature elevation into a measurable electrical signal at room temperature can enable important mid-infrared applications such as thermal sensing and imaging in ubiquitous mobile devices. However, as appealing as this goal might be, it is still unrealized due to the modest thermoelectric coefficient and weak temperature-dependence of carrier transport in graphene. Here, we demonstrate mid-infrared graphene detectors consisting of arrays of plasmonic resonators interconnected by quasi one-dimensional nanoribbons. Localized barriers associated with disorder in the nanoribbons produce a dramatic temperature dependence of carrier transport, thus enabling the electrical detection of plasmon decay in the nearby graphene resonators. We further realize a device with a subwavelength footprint of 5×5 µm$^2$ operating at 12.2 µm, an external responsivity of 16 mA/W, a low noise-equivalent power of 1.3 nW/$\sqrt{Hz}$ at room temperature, and an operational frequency potentially beyond gigahertz. Importantly, our device is fabricated using large-scale graphene and possesses a simple two-terminal geometry, representing an essential step toward the realization of on-chip graphene mid-infrared detector arrays.

**Introduction**

Surface plasmons are collective electron oscillations in conducting materials, capable of producing large optical field confinement and enhancement[1, 2]. Following optical excitation, highly confined plasmons decay inelastically within a few femtoseconds (fs), giving rise to hot charge carriers[3-5]. The carriers in turn thermalize within tens of fs at an elevated temperature $T_e$, eventually relaxing within a few picoseconds (ps) to lattice vibrations at a lower temperature $T_l$, and finally evolving toward ambient temperature $T_0$ via heat dissipation through the surrounding materials. Efficiently converting plasmons and their decay into measurable electrical signals is important to overcome the spectral limitations imposed by the bandgap of the semiconductors available for light detection and energy harvesting. Direct plasmon-to-electron conversion has been previously explored by extracting a signal along the described dissipation pipeline (e.g., via internal photoemission over a Schottky barrier[6-9]), with poor efficiency for infrared and terahertz (THz) radiation[10, 11] because of the vanishingly small plasmon energy. A radically different approach consists in monitoring the rise in electron temperature---a method that is promising when the plasmons are sustained by a comparatively small number of charge carriers (e.g., in graphene), so that a larger temperature increase is produced for a given amount of absorbed photon energy[12, 13].

Graphene has recently emerged as a promising platform for mid-infrared (mid-IR) and THz plasmonics. Compared with plasmons in noble metals, mid-IR and THz plasmons in doped graphene exhibit extraordinarily strong optical confinement and long lifetimes [14-18]. Interestingly, because of the weak electron-phonon coupling in graphene, the electronic temperature can reach significantly high values above the phonon-bath background[19]. Additionally, the large in-plane thermal conductivity of this material favors hot-carrier-assisted heat transport before dissipation into the surrounding media takes place. Despite these unique properties, the efficient electrical readout of plasmonically-induced hot-carrier generation in graphene remains an outstanding challenge. A first problem is posed by the modest Seebeck coefficient (<100 µV/K) in graphene compared to conventional thermoelectric materials[20], which renders thermoelectric detection inefficient, while the requirement of spatially varying doping[20-24] adds significant device complexity and can affect the plasmon properties. A second problem relates to the weak temperature dependence of electrical transport in graphene, especially when grown by chemical vapor-phase deposition (CVD), in which impurity scattering dominates carrier transport. In this context, graphene mid-IR bolometers relying on scattering by substrate optical phonons have been primarily explored at low temperatures ($T_0$ <10 K)[25], while at room temperature these devices exhibit unsatisfactory responsivity because of the smaller substrate's phonon-temperature elevation produced by light excitation[26].

Here, we report a simple, yet efficient two-terminal device capable of detecting thermalized carriers generated by the decay of mid-infrared plasmons excited by light of 12.2 µm wavelength. Central to our device physics is the use of strong mid-IR plasmonic resonances in discrete graphene resonators combined with quasi-1D graphene nanoribbons, whose thermally activated carrier transport is substantially influenced by the plasmonic absorption. By monitoring the conductance as a function of incident light intensity, we experimentally demonstrate a device with sub-wavelength footprint (5×5 µm$^2$) offering a high room-temperature external responsivity of 16 mA/W together with a measured noise-equivalent power (NEP) of 1.3 nW/$\sqrt{Hz}$ and a theoretical limit of 460 pW/$\sqrt{Hz}$, mainly limited by Johnson noise, thermal fluctuation noise, and shot noise. Importantly, the device is fabricated on large-scale CVD graphene, which introduces a necessary density of material defects to obtain such excellent detection capabilities and simultaneously allow for scalable fabrication, rendering it a promising candidate for the development of high-resolution mid-infrared cameras and high-density integrated infrared photonic circuits.

**Results**

Our plasmonic graphene photodetector is sketched in Fig 1a. It consists of two metal electrodes and a photoactive channel. The lithographically patterned channel is composed of multiple graphene-disk plasmonic resonators (GDPRs) connected by quasi-1D graphene nanoribbons (GNRs). GDPRs serve as both sources of thermalized carriers and leads for the GNRs. Because of the large mismatch between the geometrical dimensions of GDPRs and GNRs, the conductance of the device is dominated by that of GNRs, which host thermal carriers with appealing transport properties. First, lateral confinement produces a bandgap $\sim \theta/W$, where $\theta \sim 0.5$ eV·nm and $W$ is the GNR width [27, 28]; but more importantly, the lithographic process generates significant edge roughness in the GNRs, which in turn introduces disordered localized states, whose relative role in transport is large for quasi-1D ribbons. Carrier transport in GNRs is then affected by a dense series of disorder potentials associated with both edge roughness and charge impurities[27-30], which produce localization of the carrier wave functions. For low or moderate $T_e$, carrier transport proceeds by *hopping* between neighboring localized states [nearest-neighbor hopping (NNH)] [28] with a characteristic thermal activation energy $k_B T_{NNH}$. For sufficiently high $T_e$, thermal carrier excitation (TCE) directly over the potential barriers provides an additional transport mechanism[27, [28, 31] characterized by a larger activation energy $k_B T_{TCE}$. Here, $k_B$ is the Boltzmann constant, and $T_{NNH}$ and $T_{TCE}$ are effective temperatures characterizing NNH and TCE activation energies, respectively. Both NNH and TCE transport mechanisms are sensitive to $T_e$. Figures 1b and 1c illustrate the operation principle of our device based on these concepts. At room temperature, we have a small thermal smearing of the carrier distribution within an energy range $\sim 2k_B T_e \approx 0.05$ eV around the macroscopic chemical potential (Fig. 1b). Plasmon excitation and subsequent decay then increase the electron temperature $T_e$, resulting in larger thermal smearing of carrier energies that facilitate their passage through the landscape of disorder potentials [29, 30] (Fig. 1c). Large increase in $T_e$ triggers the TCE transport regime (Fig. 1d), which is more efficient than NNH transport (Fig. 1e). We identify excitation and decay of plasmons as a mechanism for thermal activation of the electrical conductivity in our device.

We fabricate graphene devices by dry transferring three highly doped CVD graphene layers on a 60 nm diamond-like carbon (DLC) thin film grown on a Si substrate (see Supplementary Information, SI). Note that plasmons in stacked graphene have higher frequency and larger spectral weight compared with monolayer graphene[32]. The doping of stacked graphene is around $1.5 \times 10^{13}$ cm$^{-2}$ ($E_F \sim$ -0.45 eV) as reported in prior work[33]. Compared to SiO$_2$ and other widely used substrates, the DLC thin film has a high phonon energy (165 meV), a lower surface trap density due to its nonpolar and chemically inert nature[34], and a low thermal conductivity of 0.15 W m$^{-1}$ K$^{-1}$ [35]. Figure 1f shows a microscope image of a typical device (left) and a scanning electron microscope image of the graphene nanostructures in false color (right). The GDPR diameter is 210 nm and the GNR length and width are 60 nm and 20 nm, respectively. Measured IR extinction spectra (i.e., $1 - t/t_0$, where $t$ and $t_0$ are the transmission with and without the graphene nanostructures) are plotted for light polarization either perpendicular (Fig. 1g) or parallel (Fig. 1h) to the GNRs. For both polarization directions, the GDPRs support fundamental dipolar plasmonic modes (insets to Fig. 1g-h) that produce a plasmonic peak absorption at a wavenumber ~820 cm$^{-1}$, in agreement with electromagnetic simulations (red-solid curve in Fig. 1g, see details in SI). We attribute the slightly lower plasmonic absorption and higher plasmon damping in Fig. 1h to the modified boundary condition for the dipolar mode when the incident polarization is along the GNRs.

Figure 2a (square symbols) presents the conductance $G$ of a typical device with the design shown in Fig. 1 over an ambient temperature range of 77 - 400 K, measured at a low bias voltage $V_b$= 10 mV such that Joule heating is negligible and the device operates near thermal equilibrium ($T_e = T_l = T_0$). The device has an overall size of 40×40 μm$^2$, an overall channel resistance > 5 kΩ, and a negligible contact resistance $R_c$ ~ 55 Ω (see SI). We find that the electrical conductance exhibits two distinct regimes within the measured temperature range: a NNH behavior at low temperatures, characterized by a slow increase of $G(T_e)$ with $T_e$ from 200 to 300 K; and a TCE behavior at high temperatures, when electrons acquire sufficient thermal energy to be excited and pass the disorder barriers. Recalling that $G$ is limited by GNRs, we can model its $T_e$ dependence as the sum of NNH and TCE contributions,

$$G(T_e) = B_{NNH} e^{-T_{NNH}/2T_e} N + B_{TCE} e^{-T_{TCE}/2T_e} (1-N), \qquad (1)$$

with comparable strength given by the fitting coefficients $B_{NNH}$ =186.79 μS and $B_{TCE}$ =276.17 μS. Besides the exponentials accounting for the respective fitted activation temperatures $T_{NNH}$ = 42.8 K and $T_{TCE}$ = 306.2 K, we introduce an additional $T_e$ dependence through the Fermi-Dirac-like distribution $N = \left[\exp(10T_e/T^* - 10) + 1\right]^{-1}$, in which $T^* = 342$ K acts as a characteristic temperature that separates both regimes. Equation 1 reproduces the data very well (dashed curve in Fig. 2a) and the fitted activation temperatures are in reasonable agreement with previous studies [27-30]. For fixed bias voltage (or background current $I_0$), the net current increase ($\Delta I$) in response to IR radiation (neglecting contact resistance $R_c$) can then be expressed as

$$\Delta I = I_0 \left(\frac{\Delta G / \Delta T_e}{G_0}\right) \Delta T_e, \quad (2)$$

where $G_0$ denotes the channel conductance in the absence of illumination. Hence, the quantity $(\Delta G / \Delta T)/G_0$ plotted in Fig. 2b as a function of $T_0$ provides a good measure of the device responsivity. Remarkably, it shows an increase by a factor >20 over a wide temperature range compared with measurements on an unpatterned 3-layer graphene sheet.

To further evaluate the IR photoresponse of our device, it is critical to determine the increase in electron temperature $\Delta T_e$ as a result of both Joule heating and light absorption. Here, we adopt a two-temperature model to characterize the graphene electron and lattice temperatures $T_e$ and $T_l$. Under steady-state conditions, we find that the absorbed input power ($P$) due to either electrical Joule heating or optical excitation reduces to $P = S_a A(T_e^3 - T_l^3)$, whereas conservation of heat flow imposes the condition $A(T_e^3 - T_l^3) = \kappa(T_l - T_0)$, where the cooling pathway due to carrier diffusion towards metal contacts [36, 37] is incorporated into the effective active area $S_a$ (see SI). Here, $A$ is the electron-lattice coupling coefficient, dominated by disorder-enhanced supercollision cooling [36], while $\kappa$ is the heat dissipation coefficient to the substrate. Direct solution of these two equations permits us to write the electron temperature as

$$T_{\mathrm{e}} = \left[ \left( \frac{P}{S_a \kappa} + T_0 \right)^3 + \frac{P}{S_a A} \right]^{1/3}, \qquad (3)$$

which, combined with Eqs. 1 and 2, leads to the mid-IR photoresponse shown in Fig. 2c (dashed curves). Here, we take $A$=7.89 Wm$^{-2}$K$^{-3}$ (see SI) and $\kappa$=1 MW/m$^2$K [35]. The inset of Fig. 2c shows the full solution of $T_{\mathrm{e}}$ and $T_1$, showing an electron temperature elevation $\Delta T_{\mathrm{e}} = T_{\mathrm{e}} - T_0 \sim$ 1.72 K at $T_0$ =300 K for an incident power $P_{\mathrm{inc}}$ = 660 μW. We attribute the increase in $\Delta T_{\mathrm{e}}$ at lower $T_0$ to a reduction in electron-phonon coupling and subsequent stronger hot-carrier cooling bottleneck. Note that the in-plane phonon-limited thermal conductivity of GNRs (~80 W/mK)[38] is far greater than the thermal conductivity of DLC (0.15 W/mK), which results in a negligible temperature gradient along each GNR (see finite-element-method simulations in SI), so we assume a uniform $T_{\mathrm{e}}$ over the graphene nanostructure for our analysis.

We used a quantum cascade laser (QCL) with 822 cm$^{-1}$ (12.2 μm) central emission frequency to assess the device response to mid-IR radiation. Figure 2c (symbols) shows the $T_0$ dependence of the net current increase $\Delta I(T_0)$ for two values of the QCL incident power ($P_{\mathrm{inc}}$ = 660 μW and 230 μW) and incident light polarization parallel to the GNRs. We acquired $\Delta I$ using a lock-in amplifier in series with a current preamplifier, while the bias voltage was fixed at 1 V producing a background current $I_0$ =175 μA. The results agree well with theoretical predictions (dashed curves) over a broad temperature range. We note that $\Delta I$ shows a temperature dependence similar to $(\Delta G / \Delta T)/G_0$: for $T_0$ > 300 K, we find $\Delta I$, and consequently also the external responsivity $\Delta I / P_{\mathrm{inc}}$, to increase because the device enters the TCE regime, in which the conductance shows a stronger dependence on $T_{\mathrm{e}}$. In addition to the lock-in measurements, which only reveal the absolute value of $\Delta I$, DC current-voltage measurements further validate that mid-IR radiation produces a net increase in current amplitude (see SI). This behavior is in stark contrast to the conventional bolometric effect in graphene, in which thermally-induced phonon scattering reduces the device conductance[25, 26]. Furthermore, by performing the same measurements on a device with polarization sensitive graphene plasmonic resonators (see SI), we find the contribution from graphene plasmon decay to $\Delta I$ to exceed 95%. In fact, without the plasmon resonance, Pauli blocking renders graphene with ~ -0.45 eV doping rather transparent to IR light in the investigated frequency range [39].

As an important feature of thermally activated carrier transport, the graphene device conductance is sensitive to $V_b$ due to Joule heating of the electrons. In Fig. 2d, we plot the differential conductance $dI/dV_b$ versus $V_b$ for three representative ambient temperatures ($T_0$ = 77 K, 250 K, and 360 K). To gain deeper insight into the $T_e$ change, in Fig. 2e we theoretically estimate $T_e$ and $T_l$ as functions of $V_b$ for a 40×40 μm$^2$ device. Interestingly, for $T_0$ ~ 300 K and $V_b$=1 V we find a ~5 K increase in $T_e$, which prompts us to explore Joule heating as a control knob to thermally tune the device base $T_e$ and its corresponding operation regime (i.e., TCE or NNH).

In most semiconductor-junction and quantum-well photon detectors based on absorption, when the device area ($S_d$) is smaller than the beam spot area ($S$), the photocurrent and responsivity drop significantly with decreasing $S_d$. In our device, however, upon inspection of Eq. 3, we find that plasmon-induced carrier heating $\Delta T_e$ depends on the incident light intensity instead of the total incident power. This implies that the device can maintain a similar level of $\Delta I$ and $I_0$ even if $S_d$ is scaled down, provided the sheet conductivity is the same. In fact, such behavior is accompanied by increased Joule heating resulting from the increased electrical power per unit of device area $V_b I_0 / S_d$. Under identical measurement conditions as those in Fig. 2 with $P_{inc}$ = 660 μW, we find that devices with smaller footprint areas enter the TCE regime at a much lower ambient temperature (~340 K, ~280 K, and ~250 K for 40×40 μm², 10×10 μm², and 5×5 μm² devices, respectively), as shown in Fig. 3a-c by the shadowed areas. Additionally, we observe that devices with different areas produce a change in current $\Delta I$ of the same order. This behavior is well mimicked by theory (dashed curves in Fig. 3a-c, see SI for details). In order to further clarify the role of Joule heating, we have also investigated the relationship between the photocurrent and $V_b$. For the 10×10 μm² device, at $T_0 = 77$ K under $P_{inc} = 660$ μW irradiation, the responsivity shows two distinct regimes (the two slopes in the blue data of Fig. 3d), moving from NNH to TCE with increasing $V_b$ (i.e., Joule heating power). In contrast, for the 5×5 μm² device under the same conditions, a 1V bias voltage is sufficient to drive the device into the TCE regime, and further increasing $V_b$ does not change the responsivity (red-dashed curve in Fig. 3d). It is worth noting that Joule heating plays a significant role here not only for its capability to thermally drive the device operation into the TCE regime, which offers superior responsivity, but also to augment the power dynamic range. Figure 3e presents $\Delta I$ as a function of both the total incident power $P_{inc}$ and the actual incident power on the device. The room temperature external responsivity $r_{ext} = \Delta I S / (P_{inc} S_d)$ extracted from the measured $\Delta I$ data points is found to be 16 mA/W. It is clear that the device operates more linearly as a function of incident light power when operating in the TCE regime ($T_0 = 300$ K) as compared to the NNH regime ($T_0 = 77$ K).

Remarkably, the photoresponse $\Delta I$ in our devices does not show any variation when modulating the incident light intensity over the 10 Hz $<f<$ 5 kHz frequency range (Fig. 4 inset, square dots and left axis), with the upper limit imposed by the mechanical chopper in our setup. To understand this behavior, we have performed finite-element heat transport simulations incorporating the graphene's phonon limited heat capacity and thermal conductivity, as well as its coupling to the substrate (see SI for details). The results (Fig 4 inset, red-solid curve and right axis) predict that the device remains operational at least up to >1 GHz light modulation frequencies. This bandwidth is a lower limit estimate, as it can be much larger considering the exceptionally low electronic heat capacity of graphene. We also show in Fig. 4 the measured spectral density of dark current noise ($\delta I_n$), which dramatically decreases when operated at higher frequencies with a knee at ~1 kHz. Note that 1/f noise dominates below 1 kHz, while it is strongly suppressed at higher frequency. Within the 1 kHz to 5 kHz range, there is a plateau of $\delta I_n$ with an amplitude of 21 pA/$\sqrt{Hz}$ under a bias $V_b$=1 V. Since our device can operate at speeds well beyond 1 kHz, it is not significantly affected by the large 1/f noise at low frequencies [40]. Taking the external responsivity in Fig. 3d (16 mA/W) and the measured noise amplitude into account, our device offers an experimentally derived noise-equivalent power (NEP) ~1.3 nW/$\sqrt{Hz}$. In Fig 4 we also plot the theoretical value of $\delta I_n$ corresponding to the combination of Johnson-Nyquist noise, thermal fluctuation noise, and shot noise (dashed lines), which is 7.30 pA/$\sqrt{Hz}$ under a bias $V_b$=1 V, leading to a theoretical NEP as low as 460 pW/$\sqrt{Hz}$ (see SI). We attribute the higher measured noise amplitude to contributions from the measurement circuits and amplifiers.

**Conclusions**

In summary, we have proposed and experimentally demonstrated an un-cooled device based on large-scale CVD graphene that efficiently detects mid-IR plasmon decay. The device leverages the unique physical properties of graphene, including the strong mid-IR plasmonic absorption, the small electron heat capacity, the slow hot-carrier cooling rate, the temperature-sensitive carrier transport in quasi-1D nanoribbons, and the large in-plane thermal conductivity. The electrical detection of graphene plasmons is a powerful tool for both fundamental studies and practical applications. For example, it can be used as a calorimeter to probe near-field heat transfer between graphene plasmonic structures, which is predicted to be ultrafast and highly efficient[41, 42]. As a mid-IR detector operating at room temperature, the plasmonic graphene device exhibits speed advantages compared to microbolometers [43] in mid-IR imaging and sensing. In particular, our device does not require suspended structures for thermal isolation thanks to the use of graphene, thus rendering it a viable platform for monolithic integration with other components such as optical waveguides, cavities, and electronic readout integrated circuits (ROIC). Additionally, the fast device response speed beyond gigahertz can enable applications such as free-space communications. The device concept can be readily extended into the THz spectral range by engineering the geometric and size of the GDPRs. We note that the device performance could be further improved either by reducing the doping of the quasi-1D graphene nanoribbons or by replacing them with gapped 2D semiconductors, leading to enhanced plasmonic absorption in graphene and a larger temperature dependence of the conductivity.

**References**

1.  Maier, S. A. *Plasmonics: fundamentals and applications*. Springer Science & Business Media (2007).

2.  Schuller, J. A., Barnard, E. S., Cai, W., Jun, Y. C., White, J. S., Brongersma, M. L. Plasmonics for extreme light concentration and manipulation. *Nat. Mater.* **9**, 193 (2010).

3.  Brown, A. M., Sundararaman, R., Narang, P., Goddard III, W. A., Atwater, H. A. Nonradiative plasmon decay and hot carrier dynamics: effects of phonons, surfaces, and geometry. *ACS Nano* **10**, 957-966 (2015).

4.  Narang, P., Sundararaman, R., Atwater, H. A. Plasmonic hot carrier dynamics in solid-state and chemical systems for energy conversion. *Nanophotonics* **5**, 96-111 (2016).

5.  Brongersma, M. L., Halas, N. J., Nordlander, P. Plasmon-induced hot carrier science and technology. *Nat. Nanotechnol.* **10**, 25-34 (2015).

6.  Wang, F., Melosh, N. A. Plasmonic energy collection through hot carrier extraction. *Nano Lett.* **11**, 5426-5430 (2011).

7.  Knight, M. W., Sobhani, H., Nordlander, P., Halas, N. J. Photodetection with active optical antennas. *Science* **332**, 702-704 (2011).

8.  Knight, M. W., *et al.* Embedding plasmonic nanostructure diodes enhances hot electron emission. *Nano Lett.* **13**, 1687-1692 (2013).

9.  Chalabi, H., Schoen, D., Brongersma, M. L. Hot-electron photodetection with a plasmonic nanostripe antenna. *Nano Lett.* **14**, 1374-1380 (2014).

10. Rogalski, A. *Infrared detectors*. CRC press (2010).

11. Schuck, P. J. Nanoimaging: Hot electrons go through the barrier. *Nat. Nanotechnol.* **8**, 799-800 (2013).

12. Garcia de Abajo, F. J. Graphene plasmonics: challenges and opportunities. *ACS Photon.* **1**, 135-152 (2014).

13. Yu, R., García de Abajo, F. J. Electrical detection of single graphene plasmons. *ACS Nano* **10**, 8045-8053 (2016).

14. Grigorenko, A., Polini, M., Novoselov, K. Graphene plasmonics. *Nat. Photon.* **6**, 749-758 (2012).

15. Brar, V. W., Jang, M. S., Sherrott, M., Lopez, J. J., Atwater, H. A. Highly confined tunable mid-infrared plasmonics in graphene nanoresonators. *Nano Lett.* **13**, 2541-2547 (2013).

16. Koppens, F. H., Chang, D. E., García de Abajo, F. J. Graphene plasmonics: a platform for strong light–matter interactions. *Nano Lett.* **11**, 3370-3377 (2011).

17. Ju, L., *et al.* Graphene plasmonics for tunable terahertz metamaterials. *Nat. Nanotechnol.* **6**, 630-634 (2011).

18. Yan, H., *et al.* Damping pathways of mid-infrared plasmons in graphene nanostructures. *Nat. Photon.* **7**, 394-399 (2013).

19. Gierz, I., *et al.* Snapshots of non-equilibrium Dirac carrier distributions in graphene. *Nat. Mater.* **12**, 1119-1124 (2013).

20. Hsu, A. L., *et al.* Graphene-based thermopile for thermal imaging applications. *Nano Lett.* **15**, 7211-7216 (2015).

21. Gabor, N. M., *et al.* Hot carrier–assisted intrinsic photoresponse in graphene. *Science* **334**, 648-652 (2011).

22. Song, J. C., Rudner, M. S., Marcus, C. M., Levitov, L. S. Hot carrier transport and photocurrent response in graphene. *Nano Lett.* **11**, 4688-4692 (2011).

23. Herring, P. K., *et al.* Photoresponse of an electrically tunable ambipolar graphene infrared thermocouple. *Nano Lett.* **14**, 901-907 (2014).

24. Lundeberg, M. B., *et al.* Thermoelectric detection and imaging of 1 propagating graphene plasmons. *Nat. Mater.* **16**, 204-207 (2016).

25. Yan, J., *et al.* Dual-gated bilayer graphene hot-electron bolometer. *Nat. Nanotechnol.* **7**, 472-478 (2012).

26. Freitag, M., Low, T., Zhu, W., Yan, H., Xia, F., Avouris, P. Photocurrent in graphene harnessed by tunable intrinsic plasmons. *Nat. Commun.* **4**, 1951 (2013).

27. Han, M. Y., Özyilmaz, B., Zhang, Y., Kim, P. Energy band-gap engineering of graphene nanoribbons. *Phys. Rev. Lett.* **98**, 206805 (2007).

28. Han, M. Y., Brant, J. C., Kim, P. Electron transport in disordered graphene nanoribbons. *Phys. Rev. Lett.* **104**, 056801 (2010).

29. Gallagher, P., Todd, K., Goldhaber-Gordon, D. Disorder-induced gap behavior in graphene nanoribbons. *Phys. Rev. B* **81**, 115409 (2010).

30. Stampfer, C., Güttinger, J., Hellmüller, S., Molitor, F., Ensslin, K., Ihn, T. Energy gaps in etched graphene nanoribbons. *Phys. Rev. Lett.* **102**, 056403 (2009).

31. Chen, Z., Lin, Y.-M., Rooks, M. J., Avouris, P. Graphene nano-ribbon electronics. *Physica E: Low-dimensional Systems and Nanostructures* **40**, 228-232 (2007).

32. Rodrigo, D., Tittl, A., Limaj, O., de Abajo, F. J. G., Pruneri, V., Altug, H. Double-layer graphene for enhanced tunable infrared plasmonics. *Light: Science & Applications* **6**, e16277 (2017).

33. Yan, H., *et al.* Tunable infrared plasmonic devices using graphene/insulator stacks. *Nat. Nanotechnol.* **7**, 330-334 (2012).

34. Robertson, J. Diamond-like amorphous carbon. *Materials Science and Engineering: R: Reports* **37**, 129-281 (2002).

35. Shamsa, M., Liu, W., Balandin, A., Casiraghi, C., Milne, W., Ferrari, A. Thermal conductivity of diamond-like carbon films. *Appl. Phys. Lett.* **89**, 161921 (2006).

36. McKitterick, C. B., Prober, D. E., Rooks, M. J. Electron-phonon cooling in large monolayer graphene devices. *Phys. Rev. B* **93**, 075410 (2016).

37. Crossno, J., *et al.* Observation of the Dirac fluid and the breakdown of the Wiedemann-Franz law in graphene. *Science* **351**, 1058-1061 (2016).

38. Pop, E., Varshney, V., Roy, A. K. Thermal properties of graphene: Fundamentals and applications. *MRS Bull.* **37**, 1273-1281 (2012).

39. Yan, H., *et al.* Infrared spectroscopy of wafer-scale graphene. *ACS Nano* **5**, 9854-9860 (2011).

40. Balandin, A. A. Low-frequency 1/f noise in graphene devices. *Nat. Nanotechnol.* **8**, 549-555 (2013).

41. Ilic, O., Jablan, M., Joannopoulos, J. D., Celanovic, I., Buljan, H., Soljačić, M. Near-field thermal radiation transfer controlled by plasmons in graphene. *Phys. Rev. B* **85**, 155422 (2012).

42. Yu, R., Manjavacas, A., de Abajo, F. J. G. Ultrafast radiative heat transfer. *Nat. Commun.* **8**, 2 (2017).

43. Rogalski, A., Martyniuk, P., Kopytko, M. Challenges of small-pixel infrared detectors: a review. *Rep. Prog. Phys.* **79**, 046501 (2016).

**Acknowledgments**

We acknowledge the National Science Foundation (CAREER Award) and the Office of Naval Research for the financial support. We thank Dr. X. Li for providing monolayer graphene on copper and IBM research for providing DLC on silicon substrates. F.J.G.A. and R.Y. acknowledge support from the Spanish MINECO (MAT2014-59096-P and SEV2015-0522), the European Commission (Graphene Flagship 696656), and Fundació Privada Cellex.

**Author Contributions**

Q.G. and R.Y. contributed equally to this work. Q.G. and F.X. conceived the project. Q.G. fabricated the devices and performed the measurements with help from C.L. S.Y. and B.D. Theoretical modelling and data analysis were carried out by R.Y. under the supervision of F.J.G.A. All the authors discussed the results and commented on the manuscript.

**Competing financial interests**

The authors declare no competing financial interests.

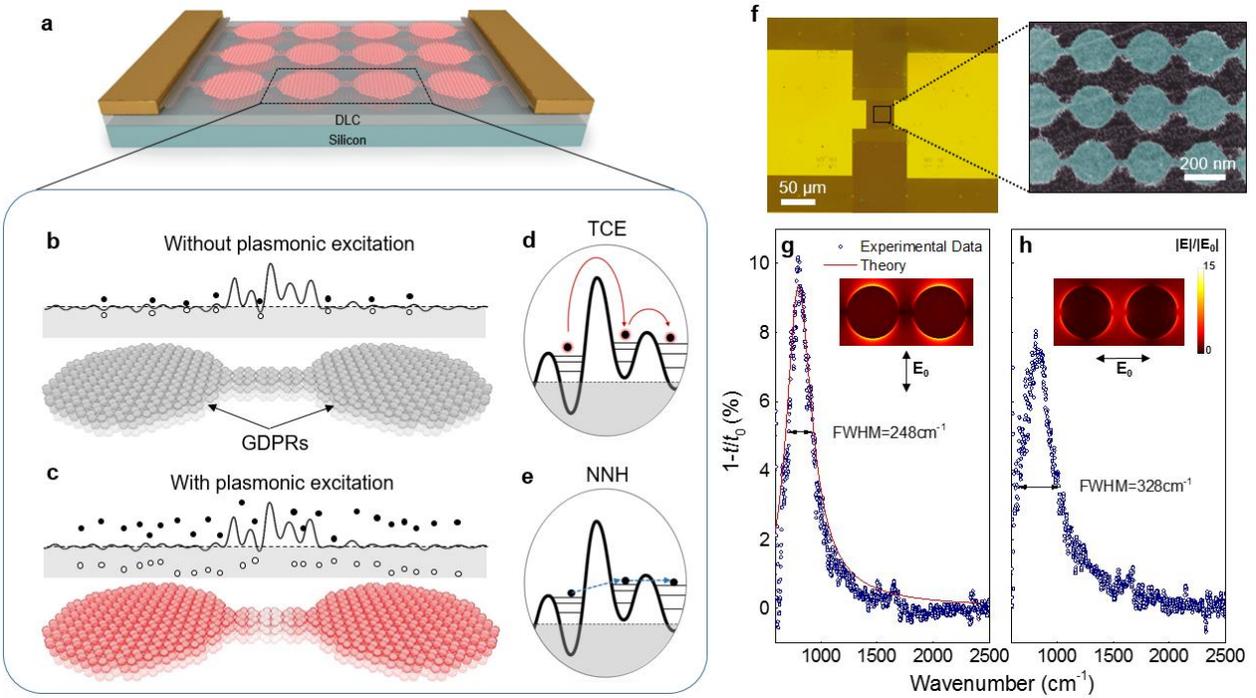

**Figure 1 | Device design and operation principle.** (a) Schematic of the proposed device, composed of graphene-disk plasmonic resonators (GDPRs, red circles) connected by quasi-1D graphene nanoribbons (GNRs). (b) Cartoon illustrating the disorder potential (solid curve) around the chemical potential before photoexcitation. The grey shadowed area denotes the states occupied by electrons, whereas filled and open circles refer to electrons and holes associated with thermal smearing at room temperature. (c) After photoexcitation on resonance with the graphene plasmons, electron-hole pairs are produced, resulting in a higher charge-carrier temperature $T_e$. (d) Illustration of thermal-carrier excitation (TCE) transport, in which electrons with higher thermal energy can overcome the localized potential barriers. (e) Illustration of nearest-neighbor hopping (NNH) transport, in which thermalized electrons evanescently hop between neighboring localized states under the driving external electric field. We use electrons to illustrate the principles of carrier transport for conceptual simplicity. (f) Optical image of the device (left) and false-color scanning electron micrograph of the graphene region (right). (g, h) Infrared extinction spectra $(1-t/t_0)$ of
17

the graphene area for incident light polarization perpendicular (g) and parallel (h) to the GNRs. Insets to (g) and (h) show simulated electric-field distributions ($|\mathbf{E}|/|\mathbf{E_0}|$) at the corresponding plasmon resonance. The solid-red curve in (g) is the calculated graphene absorption for a disk array assuming a chemical potential of -0.45 eV and a hole mobility of 475 cm$^2$V$^{-1}$s$^{-1}$.

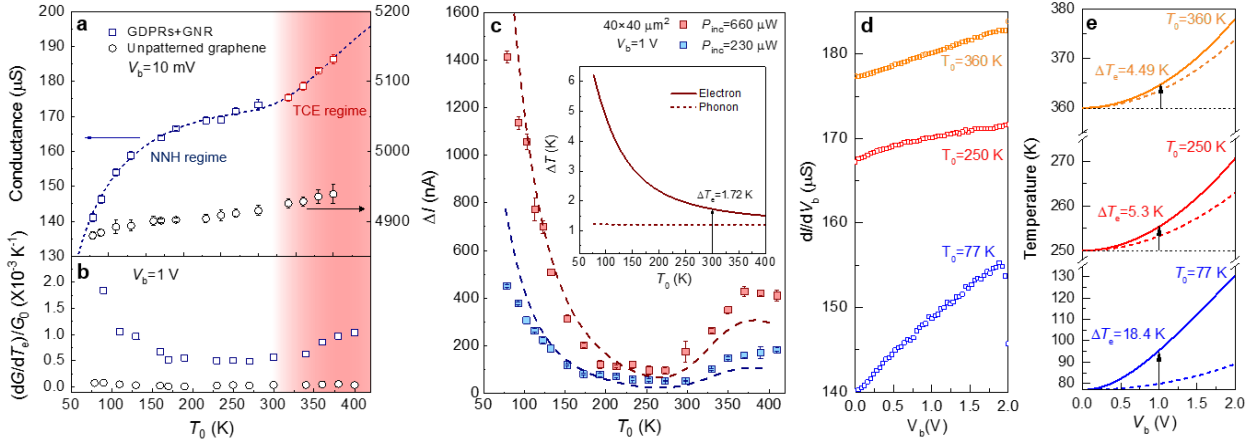

**Figure 2 | Temperature dependence of carrier transport and photocurrent generation.** (a) Conductance ($G$) versus environment temperature ($T_0$) for our graphene-plasmon device (square symbols) compared with unpatterned graphene (circles). We use 3-layer graphene and a device of 40×40 μm$^2$ area in both cases. Data are acquired with a bias voltage $V_b$=10 mV using a four-point probe configuration that eliminates the effect of contact resistance. The blue-dashed curve is a theoretical fit from Eq. 1. (b) Measured $(\Delta G/\Delta T)/G_0$ for both the graphene-plasmonic and unpatterned-graphene devices with $V_b$=1 V. (c) $T_0$ dependence of the photocurrent in the plasmonic device under 12.2 μm excitation with incident power $P_{inc}$=230 μW (blue) and 660 μW (red). Inset: calculated electron (solid curve) and phonon (dashed curve) temperature increases ($\Delta T_e$ and $\Delta T_l$) under $P_{inc}$=660 μW. The incident light is polarized parallel to the GNRs. (d) d$I$/d$V_b$ versus bias voltage $V_b$ measured at three representative environment temperatures ($T_0$=77 K, 250 K, and 360 K). (e) Calculated $T_e$ and $T_l$ as functions of $V_b$ for $T_0$ = 77 K, 250 K, and 360 K.

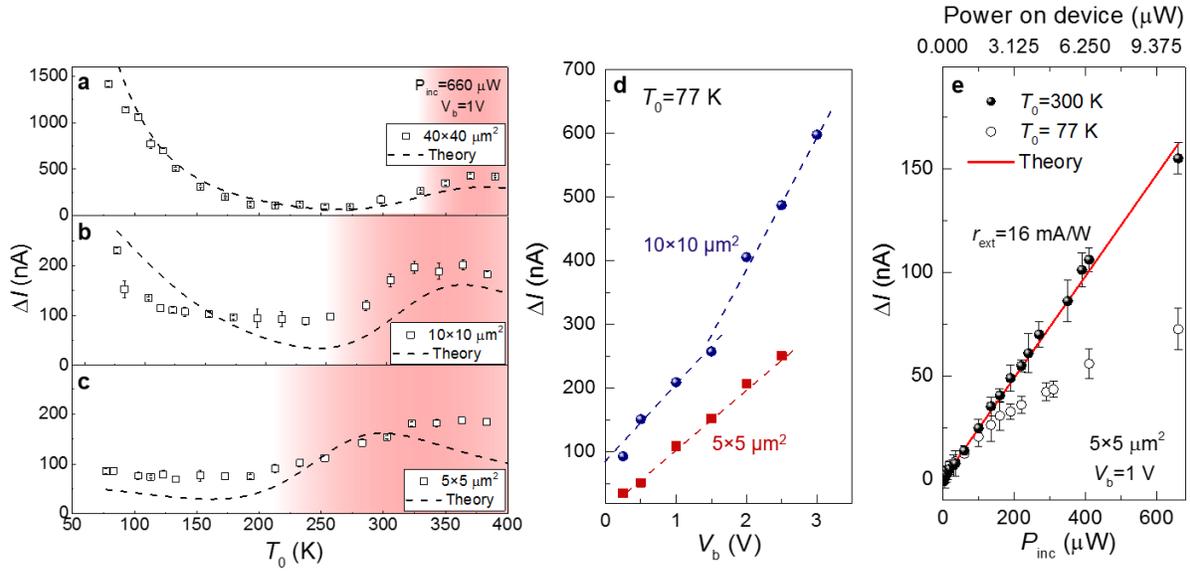

**Figure 3 | Device scalability and effect of Joule electron heating on the responsivity.** (a-c) Dependence of the photocurrent $\Delta I$ on environment temperature for devices of (A) 40×40 μm$^2$, (b) 10×10 μm$^2$, and (c) 5×5 μm$^2$ area using fixed incidence power $P_{inc}$=660 μW, bias voltage $V_b$=1 V, and light polarization (parallel to the GNRs). (d) Dependence of $\Delta I$ on $V_b$ for the 10×10 μm$^2$ (circles) and 5×5 μm$^2$ (squares) devices at an environment temperature $T_0$ = 77 K. Dashed lines are guides to the eye. (e) $\Delta I$ as a function of $P_{inc}$ for the 5×5 μm$^2$ device. The upper horizontal scale shows the power actually impinging on the device area. Filled (open) symbols represent data acquired at $T_0$=300 K (77 K). The red-solid curve is theory.

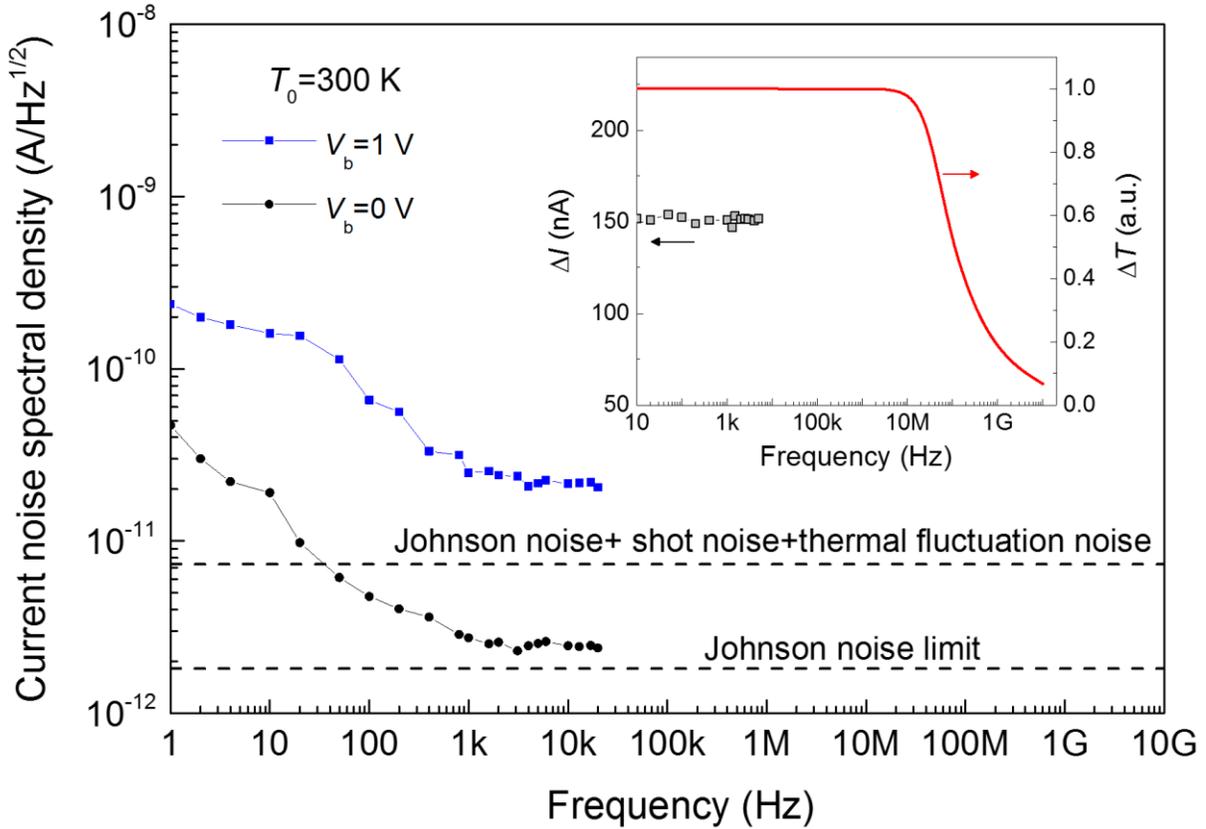

**Figure 4 | Frequency response and noise characteristics.** Spectral density of dark current noise versus frequency for zero bias (black symbols) and $V_b$=1 V (blue symbols). Each data point represents an average over 10 measurements performed at ambient temperature ($T_0$=300 K). Dashed lines show theoretical noise limits for comparison. Inset: measured photocurrent amplitude (grey squares, left axis) versus modulation frequency (10 Hz to 5 kHz) compared with the simulated frequency response (red curve, right axis) of the temperature variation in GNR regions.